# Storage Area Network Implementation on an Educational Institute Network Computer Networking and Communication

Dr. Safarini Osama
IT Department, University of Tabuk
Tabuk, KSA
usama.safarini@gmail.com, osafarini@ut.edu.sa

*Abstract* — the storage infrastructure is the foundation on which information relies and therefore must support a company's business objectives and business model.

In this environment, simply deploying more and faster storage devices is not enough; a new kind of infrastructure is needed, one that provides more enhanced network availability, data accessibility, and system manageability than is provided by today's infrastructure. The SAN meets this challenge.

The SAN liberates the storage device, so it is not on a particular server bus, and attaches it directly to the network. In other words, storage is externalized and functionally distributed across the organization. The SAN also enables the centralizing of storage devices and the clustering of servers, which makes for easier and less expensive administration. **So the idea is to** create an intelligent SAN infrastructure that stretches to meet increased demands, allows highly available and heterogeneous access to expanding information.

*Keywords-:* SAN - Storage Area Network, RAID - Redundant Array of Independent Disks, FAStT600 - Fibre Array Storage Technology, ESCON - Enterprise Systems Connection.

## I. INTRODUCTION

Networks are spread all over the world, satisfying the needs of enterprises and companies, by offering facilities of using computers and networks to meet their purposes and functions. Many challenges face these networks, which make it difficult to handle and manage the requirements of any company.

SAN is a development of networks in areas of speed, expandability, flexibility, storage, portability etc… it is like a link between present networks and future networks where we will have faster and more reliable networks. In the present time SAN is a heart of a network that we have, it can be connected to a LAN or WAN and improve the network that it is added to.

In Other Words a SAN is a centrally managed, high-speed storage network consisting of multi-vendor storage systems, storage management software, application servers and network hardware that allows you to exploit the value of your business information [1].

Storage Area Network comes out to give us many advantages on the traditional networks, such as [2]:

- Flexibility in management: so we can add or remove servers without affecting data, storage can be easily increased, changed or re-assigned, and back up servers can access a common storage pool.
- Availability: properly designed SAN storage is always available, which make it possible for many servers to access the same data pool with the same availability.
- Efficiency is provided by SAN storage by supporting a large number of operating systems and servers.
- High-speed is the ultimate goal in any network. SAN provide a very high speed by reducing the application response time, improving the processing throughput, support high performance back up, and enable new concepts as zero back up time.
- Security: as the interconnections in the SAN based on fibre channel, it will provide a very high level of security, and it supports the latest storage security measures.
- Management: it is the most important feature in SAN, as we can manage the SAN easily and from any point in the network or outside it, such as using web-based tools from any location.

These features are complementary and cumulative; that is any SAN design can use all these features, or can start with a simple SAN design then can change, add, remove or re-assign any feature we want in the future [3].

As mentioned earlier that there are many challenges faced by the network, such as protection and communication between different platforms of operating systems. With SAN we recognize the split between presentation, processing and data storage.

SAN is used to perform traditional network bottlenecks. It supports high-speed data transfers between servers and storage in the following three ways:

1- Server to storage: storage devices can be accessed serially or concurrently.
2- Server to server: for high speed, high volume communication between servers.
3- Storage to storage: outer transfer of data is done without the need of interference the work of the server.

## II. STORAGE AREA NETWORK (SAN)

A SAN is a dedicated high performance network to move data between heterogeneous servers and storage resources. A Fibre Channel based SAN combines the high performance of an I/O channel and the advantages of a network (connectivity and distance of a network) using similar network technology components like routers, switches and gateways. As shown in figure 1.

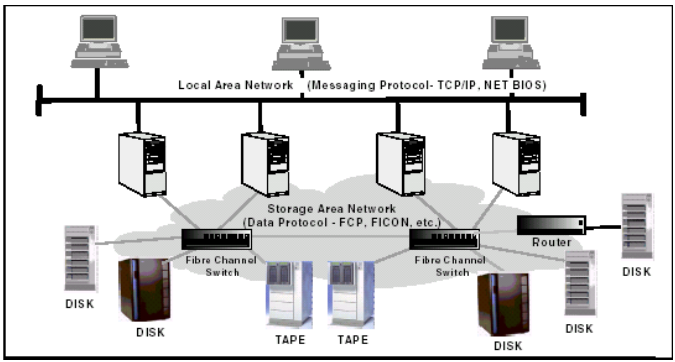

Figure 1. Connectivity and distance of a network

### Storage Area Network Environment

Storage subsystems, storage devices, and server systems can be attached to a Fibre Channel Storage Area Network. Depending on the implementation, several different components can be used to build a Storage Area Network [4]. A Fibre Channel network may be composed of many different types of interconnect entities, including switches, hubs, and bridges.

### SAN Components

**SAN components are illustrated in figure (2):**

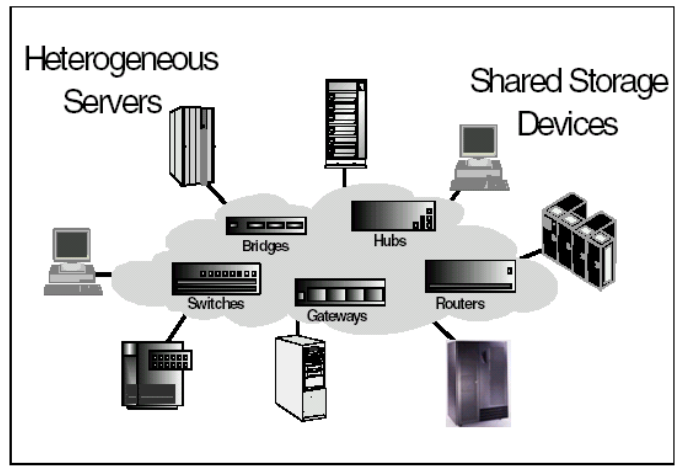

Figure 2. SAN Components

### SAN Servers

The server Infrastructure is the underlying reason for all SAN solutions. This infrastructure includes a mix of server platforms such as Windows NT, UNIX (various flavors) and OS/390. With initiatives such as Server Consolidation and e-Business, the need for SAN will increase. Although the early SAN solutions only supported homogeneous environments (e.g. future Educational institute e-mail account servers for students and e-learning based on widows NT), SAN will evolve into a truly heterogeneous environment [5].

In the last few years we were in the early SAN technology life cycle. The technology is still relatively immature and there is also a lack of accepted industrial standards. Data integrity is an important issue. Today, many current implementations rely on the physical wiring only for security, but other security schemes are being developed [6]. For example, some vendors are developing a logical partitioning approach called zoning.

Most of the SAN solutions on the market today are limited in scope and restricted to specific applications. Interoperability is not possible in many of these solutions [7]. Some of the currently available solutions are:

• Backup and recovery
• Storage consolidation
• Data replication
• Virtual storage resources

Most of the SAN management packages available today are limited in scope to the management of one particular element or aspect of the SAN. Most of these solutions are proprietary and are lacking in interoperability.

## III. IMPLEMENTATION OF SAN

The first stage is introducing the two networks that the educational institute

Owns: the Hub-based and the switch-based networks to be discussed, and then introduce the implementation of new proposed network SAN.

The Second stage is to perform a survey, which specifies the infrastructure of the proposed SAN, depending on the following factors:

I. Homogeneous environments: is based on the fact that has the same Operating Systems (Platforms), and the same vendor (maker) of the SAN.
II. Heterogeneous environment: is based on the fact that has different Operating Systems (platforms), and different vendor (maker) of the SAN.

After doing this survey, we can supply the client with the most COST EFFICTIVE SOLUTION

### Hub-based network

The Hub-based network (old Network) as shown in the figure (3) below:

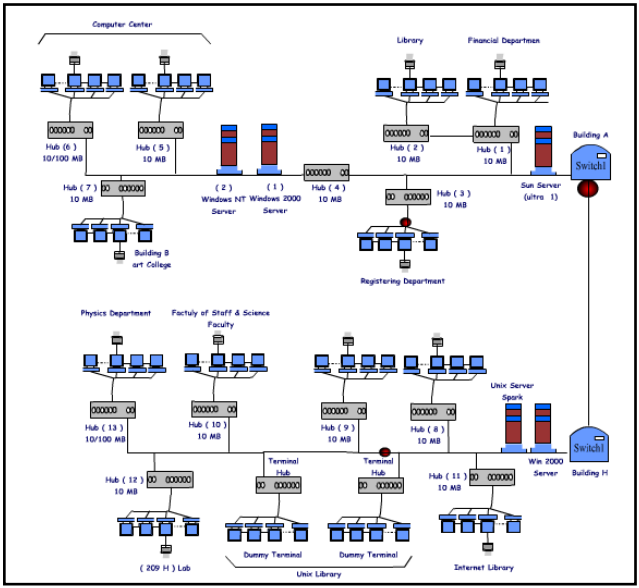

Figure 3. hub-based network

It is a weak, slow network, and is featured as one-user controls at a time, besides broadcasting, management is not easy and maintenance is also so difficult, as it is using the old wiring type.

### Switch-based network

On the other hand the switch-based network (The New LAN) is greater than the old one (Hub-based), it is shown in figures (4) and (5).

The main properties are high speed, with more Security and easy to manage than the old one. Many users can work together at the same time, besides it easier to maintain any failure in the network.

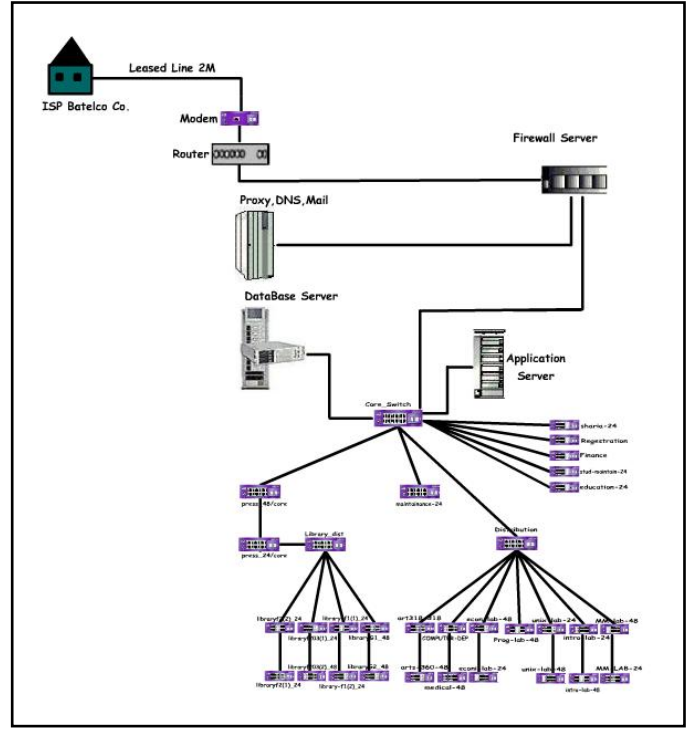

Figure 4. switch-based network (attached to the servers)

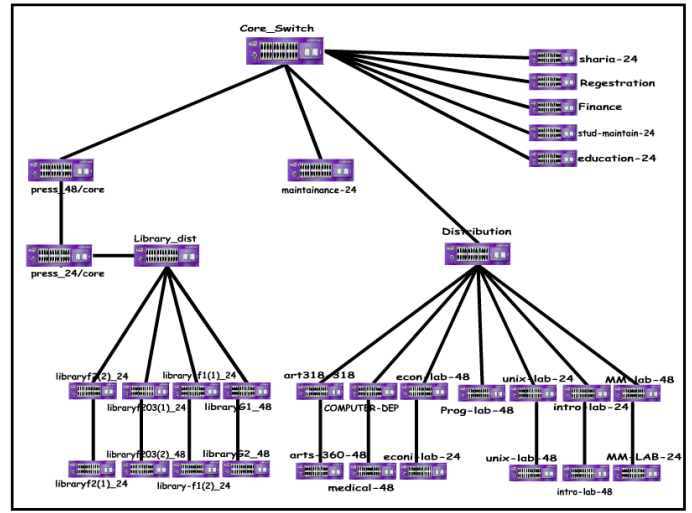

Figure 5. switch-based network

### The suggested SAN for Our Case Study (an Educational institute)

Finally is suggested the inclusion of the SAN technique for the educational institute. This will make it more powerful than the previously mentioned networks. The implementation of SAN on the institute network is shown in figures 6 and 7 below. This improvement will give the following advantages for the Educational institute network services.

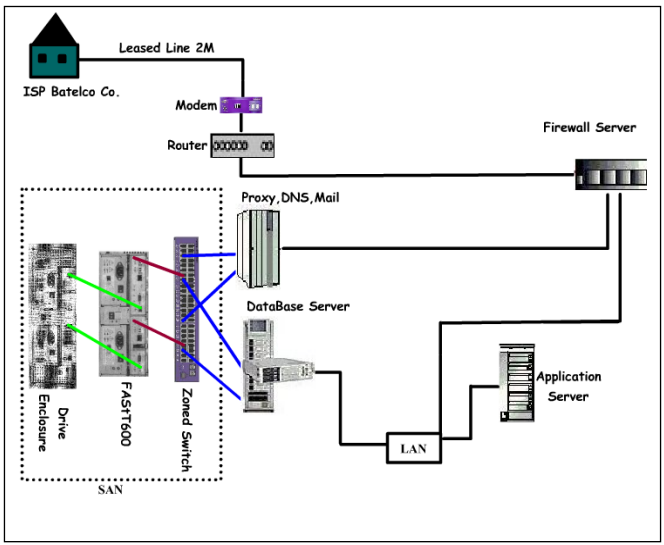

Figure 6. SAN implementation in Educational Institute (attached to the servers)

- *FAStT600* (IBM FAStT600 Storage Server)

It the cheapest storage area network and supported all requirement, that we can consider the fastt600 is the storage area network device; it containing the following equipment:

- *RAID*

There are two sample of RAID, [RAID3&RAID5] every one having a private architecture, and also perform backup recovery among all hard disks by RAID controller.

- *Hard disk*

Its storage device in SAN, they using a SCSI technology among them, and we can connect 28 hard disks in storage device, and RAID controller managing all tasks among them.

- *power supply*

It support s a storage device in a power 220V-240V OR 110V-120V.

- *Fibre channel*

It a fibre cable that consist of ESCON, it is connecting among all equipment in SAN.

- *HBA(host bus adapter)*

It's a card setting on server, using to connect between servers that having data to SAN storage by a fibre cable, and it also using a SCSI technology.

- *SCSI* *(Small Computer System Interface)*

It's a development technology that having a high performance and reliability, this technology specially used in servers because it enables to connect a large number of user with a high speed to access the server.

- *Drive Enclosure*

Is essentially a specialized chassis designed to hold and power disk drives while providing a mechanism to allow them to communicate to one or more separate computers, Drive enclosures provide power to the drives therein and convert the data sent across their native data bus into a format usable by an external connection on the computer to which it is connected.

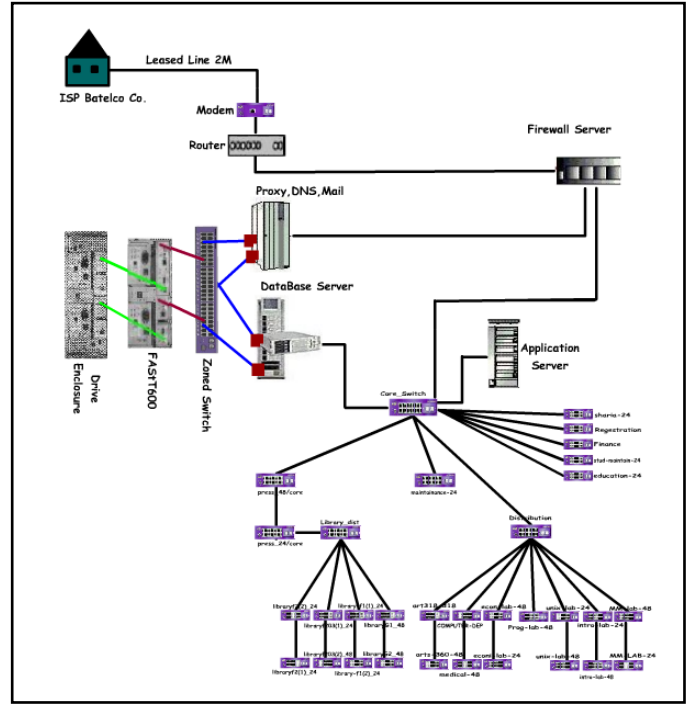

Figure 7. SAN Network implementation

IV. CONCLUSIONS

Advantages of Implementing SAN on the Network of Educational institute can be summarized as follows:

SAN can improve the speed of the network, reduces errors, make the management much more effective, it makes the backup efficient, security is greatly improved, and raises the bandwidth into a greater amount. More security involved, especially when connecting the SAN from a far distance (remote connection). Increase number of users to access server and data. Server is processed among user demands and data in SAN only, also increases performance of the server for other tasks, since the SAN removes some of the burden of the server. Backup and recovery is improved greatly in SAN, Storage consolidation in database server and mail-internet server.

**Advantages of Hub Network:**

The network was easily implemented with hubs. It was simple and inexpensive. A group of computers could be connected to the internet. Software could be shared among users. A single hardware could be connected to all computers.

**Disadvantages of Hub Network:**

Only one port works at a time so users are waiting to get control for a long time. The speed depends on number of users on the network if the number of users increases the speed goes down. Low bandwidth (data carrying capacity), the network maximum speed is 10 mbps. It depends on broadcasting for transmission so data are sending to all connected devices. There was no management so when users send data it is not secured and it is slow. Hub distribution is not organized to one another so that slows the network when sending data. Data must check every computer until it finds the target. Not easy to maintain the failure of any device if it occurs there is no way to find out. All of the writing types like (twisted pair or coaxial) have low speed. Internet maximum speed on this design is only 64 kbps.

**The Switch based Network**

The newly designed network in the other hand is much more better than the hub based network it is much faster more reliable and can be managed it also uses fibre optics which reduces the transmission error and increases bandwidth.

**Advantages of the new design (switch based network):**

Switch is smarter than hub; it reduces traffic on the network, which is a good factor for speeding the network. Security and management are factors available in switch configuration. Many users can work together over the network at the same time, with larger bandwidth. Provide greater transmission rate that reaches 100 mbps. When any device fails at any time it can be easily known and be maintained. Internet speed reaches to 10 mbps.

**Disadvantages of the new design (switch based network):**

The only disadvantage in this network is that it is much more expensive than the hub network.

SAN is the development of the network in particular aspects of speed, flexibility and storage. So similar is the link between existing networks and future networks, where networks will get faster, more reliable, but at the moment SAN is the heart of the existing networks, SAN can be connected to any network, such as (WAN, LAN), so we recommend continuous developing of SAN to become more effective.

AUTHORS PROFILE

Dr. Osama Ahmad Salim Safarini had finished his PhD. from The Russian State University of Oil and Gaz Named after J. M. Gudkin, Moscow, 2000, at a Computerized-Control Systems Department.

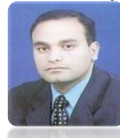

He obtained his BSC and MSC in Engineering and Computing Science from Odessa Polytechnic National State University in Ukraine 1996. He worked in different universities and countries . His research is concentrated on Automation in different branches.